\documentclass[%
reprint,
amsmath,amssymb,
aps,
]{revtex4-2}

\usepackage{graphicx}
\usepackage{dcolumn}
\usepackage{bm}

\usepackage{subcaption}
\usepackage{media9}

\usepackage{soul}

\usepackage{amssymb,amsmath}
\usepackage{algorithm}
\usepackage{algpseudocode}

\usepackage{hyperref}

\begin{document}

\preprint{APS/123-QED}

\title{Ballooning in Spiders using Multiple Silk Threads}

\author{Charbel Habchi}
\email{charbel.habchi@ndu.edu.lb}
\affiliation{Notre Dame University-Louaize, Mechanical Engineering Department, 1200 Zouk Mosbeh, Lebanon.}

\author{Mohammad K. Jawed}
\email{khalidjm@seas.ucla.edu}
\affiliation{University of California, Los Angeles, Department of Mechanical and Aerospace Engineering, Los Angeles, California 90095, USA.}

\date{\today}

\begin{abstract}
In this paper, three-dimensional numerical simulations of ballooning in spiders using multiple silk threads are performed using the discrete elastic rods method. The ballooning of spiders is hypothesised to be caused by the presence of the negative electric charge of the spider silk threads and the positive electric potential field in the earth's atmosphere. The numerical model presented here is first validated against experimental data from the open literature. After which, two cases are examined, in the first it is assumed that the electric charge is uniformly distributed along the threads while in the second, the electric charge is located at the thread tip. It is shown that the normalized terminal ballooning velocity, i.e. the velocity at which the spiders balloon after they reach steady-state, decrease linearly with the normalized lift force, especially for the tip located charge case. For the uniform electric charge case, this velocity shows a slightly weaker dependence on the normalized lift force. Moreover, it is shown in both cases that the normalized terminal ballooning velocity has no dependence on the normalized elastic bending stiffness of the threads and on the normalized viscous forces. Finally, the multi-thread bending process shows a three-dimensional conical sheet. Here we show that this behavior is caused by the Coulomb repelling forces owing to the threads electric charge which leads to dispersing the threads apart and thus avoid entanglement.
\end{abstract}

\maketitle

\section{Introduction}
Ballooning is the mechanism of dispersal of wingless arachnids, mainly spiders, to which a silk thread is attached \cite{Bell2005}. Even though, spiders do not have wings to fly, two centuries ago, Charles Darwin observed hundreds of ballooning spiders landing on the HMS Beagle located 60 miles offshore \cite{Darwin1845}. This peculiar observation, at that time, was also reported earlier in the 17th and 19th centuries \cite{Martin1669,Blackwall1827} and it was commonly occurring on relatively calm days with low wind speed, below 3 m/s \cite{Richter1970, Salmon1977, Weyman2002, Darwin2005, Lee2015}.

Since these observations \cite{Martin1669,Blackwall1827}, two competing theories were associated to explain the ballooning of spiders. In the first, ballooning of spiders is associated to natural convection currents caused by the thermal gradients in the earth boundary layer. It is assumed that these rising currents create drag forces on the light spider threads which induce lift forces when they overcome the weight of the spider. This hypothesis was extensively studied by several authors \cite{Humphrey1987,Suter1991,Weyman2002,Courtney2020,Cho2021}. For instance, Zhao \textit{et al.} \cite{Zhao2017} used a fully coupled fluid-structure interaction two-dimensional numerical model with the immersed boundary method (IBM) to analyze the effect of spider mass and thread length on the ballooning dynamics. They also analyzed the effect of vortex shedding, mainly at the trailing edge of the thread, on the oscillations and deformation of the spider silk threads during ballooning. This study is based on several assumptions neglecting the thread mass and thickness as well as representing the spider by a point mass. Suter \cite{Suter1999} studied the condition of airflow on spider ballooning, and highlighted the possibility that atmospheric turbulence may affect the ballooning and dispersal of spiders. Other authors studied the effect of atmospheric turbulence on the ballooning of spiders and on the bending of the silk threads \cite{Reynolds2006,Cho2018}. For instance, Reynolds \textit{et al.} \cite{Reynolds2006} modeled the dynamics of fully elastic silk thread in isotropic and homogeneous turbulent flows. The thread is modelled by a chain of spheres attached by springs. In their study they highlight the fact that the threads are highly twisted and bent due to turbulent structures which impede the aerodynamic control of ballooning. This fact was not captured earlier by Humphrey \cite{Humphrey1987} who modeled the spider thread by rigid inextensible massless cylindrical rod aligned with the wind direction. Meanwhile, the effect of the electric charge of spider silk threads \cite{Kronenberger2015} on the thread unfolding dynamics is not included in these aforementioned studies \cite{Cho2021}. This electric charge may induce Coulomb repelling forces which can have an important role in keeping the threads apart to avoid entanglement which may explain the three-dimensional conical sheet shape of the silk threads.

In the second hypothesis, ballooning of spiders is associated to the electrostatic force caused by the interaction between the Earth's electric field and the electric charge of spider silk threads. This electrostatic buoyancy creates a lift force on the threads which may cause spiders ballooning under certain conditions \cite{Gorham2013,Sheldon2017,Morley2018}. While, the first hypothesis discussed in the previous section is extensively analyzed in the open literature, the effect of electrostatic force on ballooning of spiders is still not fully studied and it was first introduced recently by Gorham \cite{Gorham2013}. Gorham \cite{Gorham2013} developed a simplified theoretical model of a spider with a single thread showing that the electrostatic force caused by the atmospheric potential gradient and the charged threads could be responsible for spiders take-off and ballooning. For instance, it is found \cite{Gorham2013} that a single silk thread electric charge of 100 nC is needed to lift a spider weighting 1 mg under standard atmospheric electric potential over flat field of 120 V/m \cite{Gorham2013}. From experimental observation Morley and Robert \cite{Morley2018} showed that spider mechanosensory hairs can detect electric fields which in its turn triggers the ballooning behavior. This could explain why spiders prefer to balloon from prominence, such as trees, where the electric field is higher than that on flat fields \cite{Morley2018}. Recently, Morley and Gorham \cite{Morley2020} conducted experimental measurements on ballooning behavior inside a closed chamber in which they control the electric field with no significant air motion. Coupling their experimental data to a physical one-dimensional model, they estimated that the total thread charge required for ballooning is around 1.15~nC for spiders weighting 0.9 mg, i.e. 1.28~nC/mg. In their experiment, they consider \textit{Erigone} spiders on the tip of a conductive launch point subjected to electric field strength of about 1~kV/m, similar to those observed around the tips of tree branches.

Meanwhile, in all these studies, there is no investigation on multi-thread spider ballooning process neither on the effect of the electrostatic repelling force on the terminal shape of the threads and the ballooning velocity \cite{Cho2021}. Thus, in the current paper we develop a new three-dimensional numerical model including the viscous forces, weight and dimensions of the thread and spider, electrostatic lift force and repelling forces and the elastic bending force to explore the ballooning and unfolding dynamics of spider silk threads. This can help for instance in designing new types of ballooning sensors to explore the atmospheric properties \cite{ChoANR18}.

This paper is organized as follows, in section~\ref{sec::problem} we state the problem and the physical parameters such as the spider weight, electric field, silk thread charge, viscous forces and silk thread properties. In section~\ref{sec::method} we present the numerical method and governing equations for the ballooning spider. Section~\ref{sec::results} is devoted to the results and discussions and in section~\ref{sec::conclusion} we present the concluding remarks.

\section{Methodology}
\subsection{Problem Definition}
\label{sec::problem}
It is still unclear on how spiders can emit silk threads loaded with static electric charge. According to the literature, this could be done during the spinning process where the threads are rapidly loaded with the electric charge, or this could happen after the spinning process due to friction with the air flow \cite{Gorham2013, Kronenberger2015}.

\begin{figure}[ht]
\centering
\includegraphics[width=0.3\textwidth]{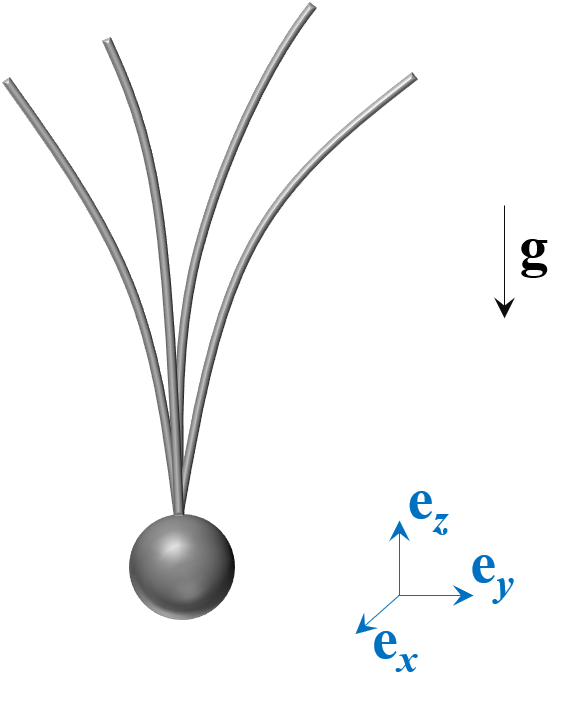}
\caption{Schematic of the spider represented by a sphere of radius $r_s$ and several threads of length $l_t$ and radius $r_t$}
\label{fig:schematic}
\end{figure}

In the present study, the spider is approximated by a sphere attached to $n_t$ silk threads initially extended vertically and very close to each other with a distance of $100~\mathrm{\mu m}$. A schematic of the spider with its threads during typical ballooning is shown in Figure~\ref{fig:schematic}.

The size and mass of the spider are chosen based on \textit{Erigone} spiders studied by Morley and Gorham \cite{Morley2018,Morley2020} where the spider mass is considered $m_s = 1~\mathrm{mg}$ and its size is $r_s = 1~\mathrm{mm}$. The typical electric charge of the spider body $Q_s$ is assumed 3~pC \cite{Morley2020}.The acceleration of gravity is $g = 9.81~\mathrm{m/s^2}$ pointing downward. The spider silk thread density is taken $\rho_t = 1200~\mathrm{kg/m^3}$ \cite{Steven1994} with a radius $r_t = 300~\mathrm{nm}$ \cite{Cho2018,Morley2020}.

The forces acting on the spider and threads are listed in this section. The weight of the spider and threads are given respectively by:

\begin{equation}
W_s = m_s g
\end{equation}

\begin{equation}
W_t = n_t \rho_t \pi r_t^2 l_t
\end{equation}
where $l_t$ is the length of one thread, $m_s$ the spider mass, $n_t$ the number of threads, $\rho_t$ the thread density and $r_t$ the thread radius.

The characteristic elastic bending force of the threads is expressed as:

\begin{equation}
E_b = \frac{YI}{l_t^2}
\end{equation}
where $Y$ is the thread's Young modulus and $I = \frac{\pi r_t^4}{4}$ is the area moment of inertia of the silk thread.

The Coulomb repulsion force acting on the threads is given by the Coulomb's inverse-square law:

\begin{equation}
F_r = k_e \frac{\left|q_1 q_2\right|}{r^2}
\end{equation}
where $k_e = 9 \times 10^9~\mathrm{N m^2 C^{-2}}$ is the Coulomb constant, $q_1$ and $q_2$ are the signed magnitudes of the charges of the spider threads, and $r$ is the distance between the threads.

The electrostatic force of the threads and the spider is given by the following expression:

\begin{equation}
F_l = E \left(Q_t + Q_s\right)
\end{equation}
where $E$ is the earth electric field, $Q_t$ and $Q_s$ are the total charges of the silk threads and spider, respectively.

Finally the hydrodynamic forces $F_v$ on the spider and threads are computed using the Resistive Force Theory (RFT) as explained in the next section.

In this paper we run parametric sweep simulations by varying the forces acting on the spider and analyzing the normalized terminal ballooning velocity, which is the normalized speed at which the spider is ballooning when it reaches steady-state. Moreover, the unfolding dynamics of the spider threads under the coupled effect of electrostatic and viscous forces are studied.

\subsection{Numerical Method for Ballooning Spiders}
\label{sec::method}
The numerical method adopted to study the fluid-structure-electric field interaction combines three components. The first component concerns the Discrete Elastic Rod (DER) method to compute the elastic deformation of the threads \cite{Bergou2010}, i.e., bending, twisting, and stretching, with the primary mode of deformation being bending. The second component is the RFT adopted to compute the hydrodynamic viscous forces on the spider and threads \cite{Lighthill1976}, and finally the third component is the electrostatic forces caused by the atmospheric potential gradient and the silk electric charge.

\begin{figure*}[ht]
\centering
\includegraphics[width=1\textwidth]{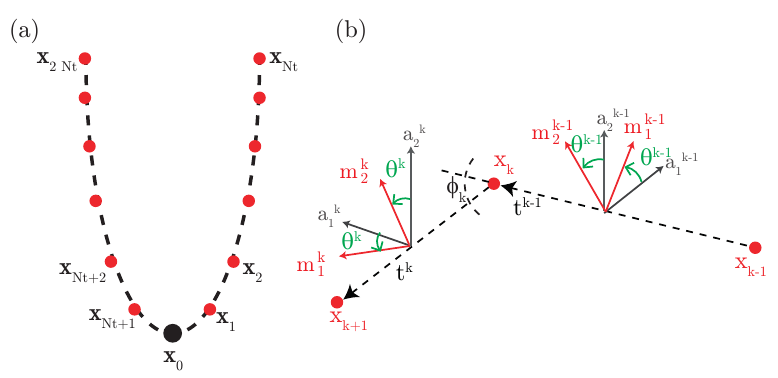}
\caption{(a) Discrete representation of a spider with $2$ threads. (b) Three nodes, two edges, and the associated material and reference frames. Bending energy is related to the turning angle $\phi_k$, twisting energy is related to $(\theta^k - \theta^{k-1})$, and stretching energy is related to the elongation of the edges.
}
\label{fig:DER_schematic}
\end{figure*}

The numerical simulations in this paper employ a discrete kinematic representation of the spider following the DER algorithm \cite{Bergou2010, bergou2008discrete, jawed2018primer}. In Figure~\ref{fig:DER_schematic}(a), the spider is modeled as a network of elastic rods with one node, $\mathbf x_0$, representing the spider body and $N_t$ nodes per thread. For a spider with $n_t$ threads, the total number of nodes is $n_t N_t + 1$. The vector between two consecutive nodes is an ``edge" and each thread is composed of $N_t$ edges. The edges, $\mathbf e$, on the $j$-th thread are

\begin{eqnarray*}
\mathbf e^{N_t (j-1) + 0} = \mathbf x_{N_t (j-1) + 1} - \mathbf x_0,\\
\mathbf e^{N_t (j-1) + 1} = \mathbf x_{N_t (j-1) + 2} - \mathbf x_{N_t (j-1) + 1},\\
\mathbf e^{N_t (j-1) + 2} = \mathbf x_{N_t (j-1) + 3} - \mathbf x_{N_t (j-1) + 2},\\
\ldots, \\
\mathbf e^{N_t j - 1} = \mathbf x_{N_t j} - \mathbf x_{N_t j-1}.
\end{eqnarray*}

Note that an edge can be usually defined as $\mathbf e^i = \mathbf x_{i+1} - \mathbf x_i$ (i.e. vector connecting two consecutively numbered nodes), except the first edge on each thread. The total number of edges for a spider with $n_t$ threads is $n_t N_t$.

In order to keep track of the rotation of the edges, the $k$-th edge in Figure~\ref{fig:DER_schematic}(b) is decorated with an orthonormal material frame, $\left( \mathbf m_1^k, \mathbf m_2^k, \mathbf t^k \right)$, where $\mathbf t_k$ is the unit normal vector parallel to $\mathbf e^k$ (i.e. tangent along the $k$-th edge). Since this frame always has the third director parallel to the tangent, it is an ``adapted" frame. A reference frame $\left( \mathbf a_1^k, \mathbf a_2^k, \mathbf t^k \right)$ -- another orthonormal adapted frame -- is also associated with each edge. At time $t=0$, the reference frame and the material frame are identical. During the time marching scheme of the simulation (Algorithm~\ref{algo:discreteElasticRods}), the reference frame is updated through parallel transport in time. Parallel transport is the {\em most natural} or {\em twist free} way of moving an adapted frame from one edge to another; details can be found in Refs.~\cite{jawed2018primer, Bergou2010}. Using the reference frame, the material frame can be fully described using a single scalar quantity -- the twist angle, $\theta^k$, -- which is the signed angle from $\mathbf a_1^k$ to $\mathbf m_1^k$ about the tangent $\mathbf t^k$.

The degrees of freedom (DOF) vector $\xi$ of the spider with $n_t N_t+1$ nodes and $n_t N_t$ edges has a size of $\mathrm{ndof} = 3 \times (n_t N_t+1) + n_t N_t$ and is defined as:
\begin{equation}
\mathbf \xi = \left[
\mathbf x_0, \mathbf x_1, \mathbf x_2, \ldots, \mathbf x_{n_t N_t}, \theta^0, \theta^1, \ldots, \theta^{n_t N_t-1} \right]^T,
\end{equation}
where the superscript $^T$ denotes transpose. The equation of motion at each DOF is
\begin{equation}
m_i \frac{\partial^2 \xi_i}{\partial t^2} + \frac{\partial E_{\textrm{elastic}}}{\partial \xi_i} -
f_i^{\textrm{ext}} = 0,
\label{eq:DER3D_EOM_Continuous}
\end{equation}
where $i=1, \ldots, \mathrm{ndof}$, $E_{\textrm{elastic}}$ is the elastic energy responsible for the stretching and bending, $f_i^{\textrm{ext}}$ is the external force (or moment for twist angles), e.g. gravity, and $m_i$ is the lumped mass at each DOF. The lumped mass at the head node which represents the spider mass is $m_s$; the mass on the other nodes is computed using the density of thread $\rho_t$, its cross-sectional radius $r_t$ and the length of the discrete edges. For the DOFs representing rotation (twist angles), the lumped mass is $\frac{1}{2} \Delta m r_t^2$, where $\Delta m$ is the mass of an edge and $r_t$ is the silk thread radius.
The simulation discretizes time into small steps and $\Delta t$ is the time step size. The equation of motion to march from $t=t_j$ to $t=t_{j+1}=t_j + \Delta t$ is
\begin{equation}
f_i \equiv \frac{m_i} {\Delta t} \left[
\frac{ \xi_i (t_{j+1}) - \xi_i (t_j) } { \Delta t } -
\dot{\xi}_i (t_j) \right] +
\frac{\partial E_{\textrm{elastic}}}{\partial \xi_i} -
f_i^{\textrm{ext}} = 0,
\label{eq:DER3D_EOM}
\end{equation}
where $f_i$ is the force exerted on each node. The {\em old} DOF $\xi_i (t_j)$ and velocity $\dot{\xi}_i (t_j)$ from the previous time step are known, $E_{\textrm{elastic}}$ is the elastic energy evaluated at $\xi_i (t_{j+1})$, and $f_i^{\textrm{ext}}$ is the external force evaluated at $\xi_i (t_{j+1})$.

In the simulations reported in this paper, number of nodes per threads is $N_t = 10^2$ and the time step size, $\Delta t$, is always less than $10^{-1}$s. We use an adaptive time stepping scheme where the time step size is automatically reduced by a factor of $10$ if the simulation fails to converge and is increased by a factor of $10$ (but always less than $10^{-1}$s) if the simulation runs successfully for approximately $10$ time steps.

The Jacobian for equation~\ref{eq:DER3D_EOM} is

\begin{equation}
\mathbb J_{ij} = \frac{\partial f_i}{\partial \xi_j} = \mathbb J^{\textrm{inertia}}_{ij} + \mathbb J^{\textrm{elastic}}_{ij} + \mathbb J^{\textrm{ext}}_{ij},
\label{eq:DER3D_Jacobian}
\end{equation}

where
\begin{eqnarray}
\mathbb J^{\textrm{inertia}}_{ij} = \frac{m_i}{\Delta t^2} \delta_{ij},\\
\mathbb J^{\textrm{elastic}}_{ij} = \frac{\partial^2 E_{\textrm{elastic}}}{\partial q_i \partial q_j},\\
\mathbb J^{\textrm{ext}}_{ij} = - \frac{\partial f_i^{\textrm{ext}}} {\partial q_j}.
\label{eq:jac_extForce}
\end{eqnarray}

Here, $\delta_{ij}$ represents Kronecker delta. We can solve the $\mathrm{ndof}$ equations of motion in equation~\ref{eq:DER3D_EOM} to obtain the {\em new} DOF $\mathbf \xi (t_{j+1})$. The new velocity is simply

\begin{equation}
\dot{ \mathbf \xi} (t_{j+1}) = \frac{\mathbf \xi (t_{j+1}) - \mathbf \xi (t_j)}{\Delta t}.
\end{equation}

Evaluation of the gradient of the elastic energy ($\frac{\partial E_{\textrm{elastic}}}{\partial \xi_i}$) as well as its Hessian ($\frac{\partial^2 E_{\textrm{elastic}}}{\partial \xi_i \partial \xi_j}$) are well documented in Refs.~\cite{jawed2018primer, Bergou2010, panetta2019x}. Bending energy is associated with the {\em turning angle} ($\phi_k$ in Figure~\ref{fig:DER_schematic}) at the {\em internal nodes} on each thread, e.g. in case of the $j$-th thread, the associated nodes are $\mathbf x_{(j-1)N_t+1}, \mathbf x_{(j-1)N_t+2}, \ldots, \mathbf x_{j\,N_t-1}$. Twisting energy is associated with the same nodes. Stretching energy is associated with each edge.

Unique to the problem of ballooning of spiders is the external forces, described next. Four types of external forces are acting on the rod network such that the $\mathrm{ndof}$-sized external force vector (cf. equation~\ref{eq:DER3D_EOM}) is
\begin{equation}
\mathbf f^{\textrm{ext}} = \mathbf W + \mathbf F_v + \mathbf F_r + \mathbf F_l,
\label{eq:externalForce}
\end{equation}
where the term $\mathbf W$ is the weight vector which can be trivially computed from the weight of the spider body, the density of the threads, and their cross-sectional radius. The viscous force term $\mathbf F_v$ exerted by the surrounding air on the $k$-th node (cf. Figure~\ref{fig:DER_schematic}(b)) to march from $t=t_j$ to $t=t_{j+1} = t_j + \Delta t$. Following Gray and Hancock\rq{}s RFT~\cite{coq2008rotational, gray1955propulsion}, the force on the node is
\begin{equation}
\mathbf F_{v, k} = \left( -\eta_\parallel + \eta_\perp \right) \mathbf t_k \, \mathbf t_k^T \Delta l \mathbf v_k - \eta_\perp \Delta l \mathbf v_k,
\label{eq:RFT}
\end{equation}
where $\Delta l$ is the Voronoi length ($\frac{l_t}{N_t}$ for the internal nodes on the thread and $\frac{l_t}{2N_t}$ for the terminal nodes), $\mathbf t_k$ is the node-based tangent (average of the tangents on the edge before and the edge after the $k$-th node), $\mathbf v_k = \frac{\mathbf x_k (t_{j+1}) - \mathbf x_k (t_j)} {\Delta t}$ is the velocity of the $k$-th node, and the Resistive Force coefficients are
\begin{eqnarray}
\eta_\parallel = \frac{2 \pi \mu}{\log \left(\frac{l_t}{r_t}\right) - \frac{1}{2}},\\
\eta_\perp = \frac{4 \pi \mu} {\log \left(\frac{l_t}{r_t}\right) + \frac{1}{2}}, \end{eqnarray}
where $\mu$ is the dynamic viscosity of air.

Note that $\eta_\perp$ is approximately twice of $\eta_\parallel$, i.e. the resistance from drag is lower when the motion is along the tangent and higher when the motion is perpendicular to the tangent. The force calculated at each node using equation~\ref{eq:RFT} is used to populate the $\mathrm{ndof}$-sized viscous force vector, $\mathbf F_{v}$.

The spider body is assumed spherical, and thus the drag force at $x_0$ is computed using Stokes law given as follows:

\begin{equation}
F_{v,0} = 6 \pi \mu r_s v_s
\end{equation}
where $r_s$ is the sphere radius and $v_s$ is the spider body speed.

It should be noted that RFT is a simplification that ignored the hydrodynamic interaction induced by distant parts of one or multiple threads. This is in contrast with more accurate slender body theories (SBT)~\cite{rodenborn2013propulsion} that capture this interaction. While recent works~\cite{Huang2021} have combined DER with SBT, RFT seems to be reasonably accurate when the rod has a low curvature~\cite{rodenborn2013propulsion}. Further, RFT can be included in the simulation using the backward Euler\rq{}s method (i.e. the gradient of the external force with respect to the DOFs in equation~(\ref{eq:jac_extForce}) is known). This is not the case when SBT is used and the SBT-derived force has to be incorporated using the forward Euler\rq{}s method. Moreover, SBT requires solving a dense linear system of size $\mathrm{ndof}$; this worsens the time complexity of the algorithm. The interaction between the flows induced by the head and the threads has also been ignored in our setup. It is possible to incorporate this interaction~\cite{thawani2018trajectory} at the expense of computational efficiency. However, in this study, the aim is to explore the essential physics of the ballooning phenomenon by parameter space exploration in numerical simulations and therefore a computationally efficient framework with DER and RFT has been chosen. A more comprehensive model is an interesting direction for future research.

The term $\mathbf F_r$ in equation~(\ref{eq:externalForce}) is the Coulomb repulsion force on the $k$-th node given as:
\begin{equation}
\mathbf F_{r, k} = k_e \sum_{i \neq 0, i \neq k} \frac{q_i q_k}{r_{i,k}^3} \mathbf r_{i,k},
\label{eq:Coulomb_Appendix}
\end{equation}
where $r_{i,k} = \| \mathbf x_i - \mathbf x_k \| $ is the Euclidean norm of the distance between two nodes, $k_e = 9 \times 10^9~\mathrm{N m^2 C^{-2}}$ is the Coulomb constant, and $q_i$ is the charge located at the $i$-th node. At the first node (spider body), the charge is $q_0 = Q_s$, where $Q_s$ is the spider body electric charge. For all the other nodes (nodes on the thread of the spider), $q_i$ can be computed from the total thread charge $Q_t$. Two cases will be discussed in Section~\ref{sec:ballooningSpeed}. In the first case, the electric charge is located at the thread tip and $q_i = Q_t$ at the tip nodes ($\mathbf x_{Nt}$ and $\mathbf x_{2Nt}$ in Fig.~\ref{fig:DER_schematic}); $q_i=0$ otherwise. In the second case, the electric charge is uniformly distributed along the thread and $q_i = Q_t \Delta l / l_t$, where $\Delta l$ is the length of each edge and $l_t$ is the length of each thread. The force calculated at each node using equation~\ref{eq:Coulomb_Appendix} then constitutes the Coulomb repulsion vector, $\mathbf F_{r}$, of size $\mathrm{ndof}$.

Finally the electrostatic lift term $\mathbf F_l$ which only acts along the $z$-axis. At the $k$-th node on the rod network, the lift force vector (size $3$) is
\begin{equation}
\mathbf F_{l, k} =
\begin{bmatrix}
0,\\
0,\\
E_k q_k
\end{bmatrix},
\label{eq:lift}
\end{equation}
where $E_k$ is the electric potential evaluated at $z$-coordinate of the node, $\mathbf x_k$, at $t=t_{j+1}$ from equation (\ref{eq::elecMorley}), and $q_k$ is the charge located at the $k$-th node. The charge located on the head node, $\mathbf x_0$, is different than the charges located on the thread nodes. After calculating the forces on each node, the $\mathrm{ndof}$-sized electrostatic lift force, $\mathbf F_l$, can be constructed.

In the force expressions above, we did not explicitly write down the Jacobian terms (e.g. derivative of the forces with respect to the DOFs). However, derivation of the Jacobian terms related to these external forces require is rather trivial.

The main steps of the algorithm are outlined below in Algorithm~\ref{algo:discreteElasticRods}.

\begin{algorithm}[H]
\caption{Discrete Elastic Rods}\label{algo:discreteElasticRods}
\begin{algorithmic}[1]
\Require{$\mathbf \xi (t_j), \dot{\mathbf \xi} (t_j)$} \Comment{DOFs and velocities at $t=t_j$}
\Require{$\left( \mathbf a_1^k (t_j), \mathbf a_2^k (t_j), \mathbf t^k (t_j) \right)$, $k \in \left[ 0, n_t N_t-1 \right]$} \Comment{Reference frame at $t=t_j$}
\Ensure{$\mathbf \xi (t_{j+1}), \dot{\mathbf \xi} (t_{j+1})$} \Comment{DOFs and velocities at $t=t_{j+1}$}
\Ensure{$\left( \mathbf a_1^k (t_{j+1}), \mathbf a_2^k (t_{j+1}), \mathbf t^k (t_{j+1}) \right)$, $k \in \left[ 0, n_t N_t-1 \right]$} \Comment{Reference frame at $t=t_{j+1}$}
\Statex
\Function{Discrete\_Elastic\_Rods}
{$\; \mathbf \xi, \dot{\mathbf \xi} (t_j), \left( \mathbf a_1^k (t_j), \mathbf a_2^k (t_j), \mathbf t^k (t_j) \right)\;$}
\State {Guess: $\mathbf \xi^{(1)} (t_{j+1}) \gets \mathbf \xi (t_{j})$}
\State {$n \gets 1$}
\While{\texttt{error} $>$ \texttt{tolerance}}
\State Compute reference frame $
\left( \mathbf a_1^k (t_{j+1}), \mathbf a_2^k (t_{j+1}), \mathbf t^k (t_{j+1}) \right)^{(n)}$ using $\mathbf \xi^{(n)} (t_{j+1})$
\State Compute reference twist $\Delta m_{k, \textrm{ref}}^{(n)}$ at each internal node
\State Compute material frame $
\left( \mathbf m_1^k (t_{j+1}), \mathbf m_2^k (t_{j+1}), \mathbf t^k (t_{j+1}) \right)^{(n)}$
\State Compute $\mathbf f$ and $\mathbb{J}$ \Comment{Equations.~\ref{eq:DER3D_EOM} and \ref{eq:DER3D_Jacobian}}
\State $\Delta \mathbf \xi \gets \mathbb J \backslash \mathbf f$ \Comment{Newton-Raphson method}
\State $\mathbf \xi^{(n+1)} \gets \mathbf \xi^{(n)} - \Delta \mathbf \xi$ \Comment{Update DOFs}
\State \texttt{error} $\gets$ \texttt{ sum ( abs ( } $\mathbf f$ \texttt{) ) }
\State $n \gets n+1$
\EndWhile
\Statex
\State $\mathbf \xi (t_{j+1}) \gets \mathbf \xi^{(n)} (t_{j+1})$
\State $\dot{\mathbf \xi} (t_{j+1}) \gets \frac{ \mathbf \xi (t_{j+1}) - \mathbf \xi (t_j) } {\Delta t}$
\State $\left( \mathbf a_1^k (t_{j+1}), \mathbf a_2^k (t_{j+1}), \mathbf t^k (t_{j+1}) \right) \gets
\left( \mathbf a_1^k (t_{j+1}), \mathbf a_2^k (t_{j+1}), \mathbf t^k (t_{j+1}) \right)^{(n)}$
\State \Return {$\mathbf \xi (t_{j+1}), \dot{\mathbf \xi} (t_{j+1}), \left( \mathbf a_1^k (t_{j+1}), \mathbf a_2^k (t_{j+1}), \mathbf t^k (t_{j+1}) \right)$}
\EndFunction
\end{algorithmic}
\end{algorithm}

\subsection{Validation for Single Thread}
\label{sec:validation}
In this section we validate the results computed numerically for a single thread case with those obtained by Gorham~\cite{Gorham2013} and Morley and Gorham~\cite{Morley2020}. In their studies, a spider with a single thread was assumed, thus, the repelling forces between the threads are not considered in their models.

In the first validation study, the electric potential field given in equation (\ref{eq::potentialGorham}) is adopted as in the theoretical analysis of Gorham~\cite{Gorham2013}. Gorham \cite{Gorham2013} used an approximated analytical model for the atmospheric electric field as it varies with the altitude from flat earth surface on a normal day:

\begin{equation}
E = E_0 e^{-\alpha z}
\label{eq::potentialGorham}
\end{equation}
where $E_0 = -120 \mathrm{Vm^{-1}}$ is the reference electric field at zero altitude, $\alpha = 3 \times 10^{-4} \mathrm{m^{-1}}$ and $z$ is the altitude in m.

\begin{figure}[ht]
\centering
\includegraphics[width=0.5\textwidth]{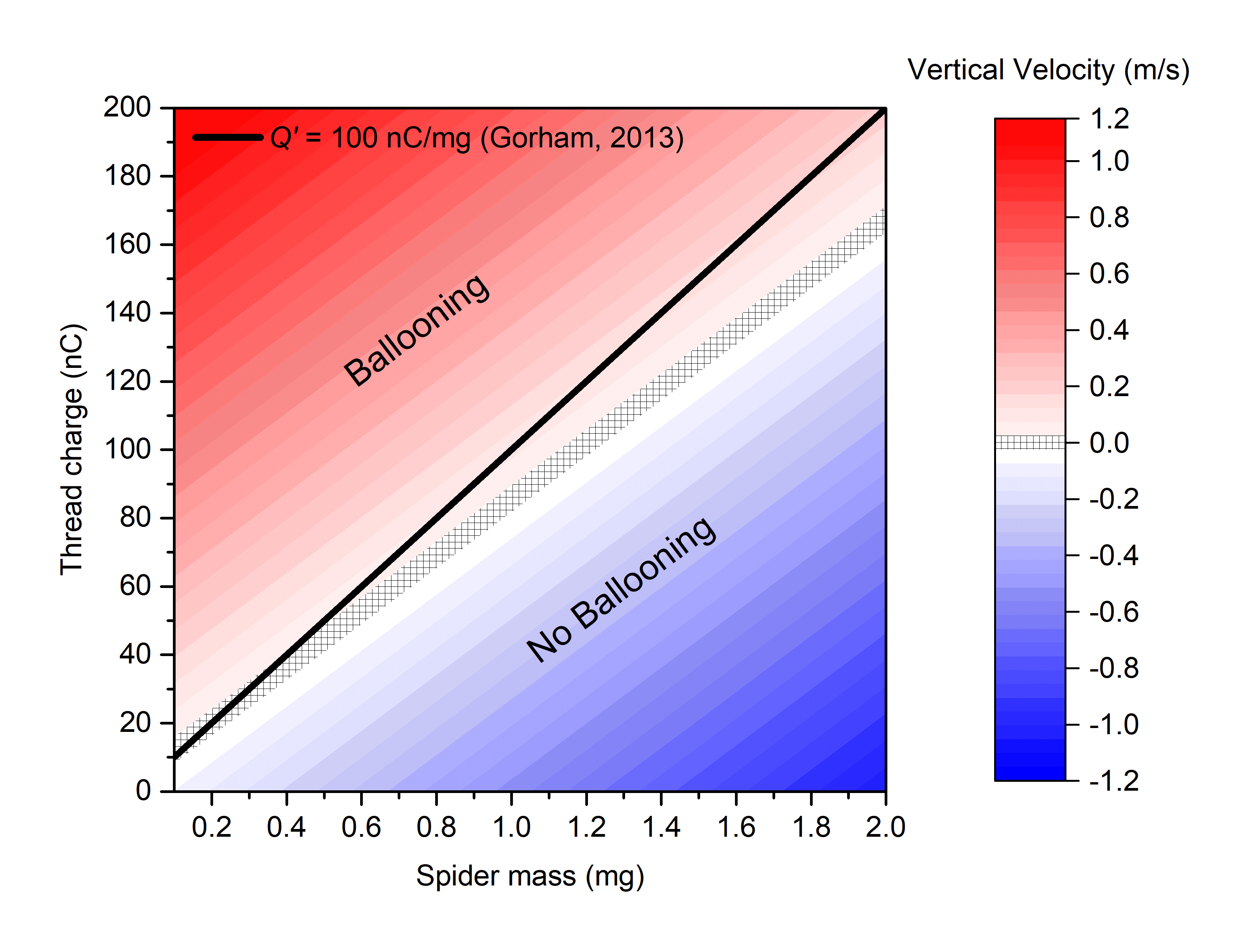}
\caption{Contour of vertical velocity versus spider mass and charge electric charge. The black solid line corresponds to the required charge for ballooning per 1 mg of spider mass obtained theoretically by Gorham~\cite{Gorham2013}. The hatched region in the plot corresponds to the 86 nC/mg of required charge for ballooning obtained from present numerical simulations.}
\label{fig:validationGorham}
\end{figure}

Simulations are carried out by spanning spiders with masses from 0.1 to 2 mg and thread electric charge from 10 to 200 nC. We observe the vertical velocity of the spider; if this velocity is positive this means that ballooning will eventually take place. If the velocity is negative it means that the spider will be in a free fall. Figure~\ref{fig:validationGorham} shows the contour of vertical velocity versus spider mass and electric charge in addition to the required thread charge to obtain ballooning found theoretically by Gorham~\cite{Gorham2013} and which corresponds to 100 nC/mg of spider mass. From this figure, it is observed that the terminal ballooning velocity reaches up to 1.2 m/s for small spiders weighting 0.1 mg and having a thread charge of 200 nC. From the present numerical simulations the required ballooning for single thread spider is around 86 nC/mg which is $14~\%$ different than that obtained by Gorham~\cite{Gorham2013} theoretical analysis which was simplified by assuming a ballooning acceleration of $3~\mathrm{m/s^2}$ and neglected viscous effects.

Thus in the second validation we consider the experimental measurement coupled to 1-D numerical simulations of ballooning spiders performed by Morley and Gorham \cite{Morley2020} inside a controlled closed chamber. In their study, the electric potential field is computed using a commercial electromagnetic simulation software. The resulting electric potential obtained in \cite{Morley2020} is fitted here using the following exponential function and implemented in our simulations:

\begin{equation}
E = E_1 \exp(-z/z_1) + E_2 \exp(-z/z_2) + E_0,
\label{eq::elecMorley}
\end{equation}
where $E_0 = 7.41 \times 10^3~\mathrm{V/m}$, $E_1 = 2.52 \times 10^5~\mathrm{V/m}$ and $z_1 = 1.51 \times 10^{-3}~\mathrm{m}, E_2 = 5.07 \times 10^4~\mathrm{V/m}$ and $z_2 = 7.93 \times 10^{-3}~\mathrm{m}$.
Following this linear regression model adopted in equation (\ref{eq::elecMorley}), the $R^2$ value is around 0.9973, which is evidence of the good fitting.

\begin{figure}[ht]
\centering
\includegraphics[width=0.5\textwidth]{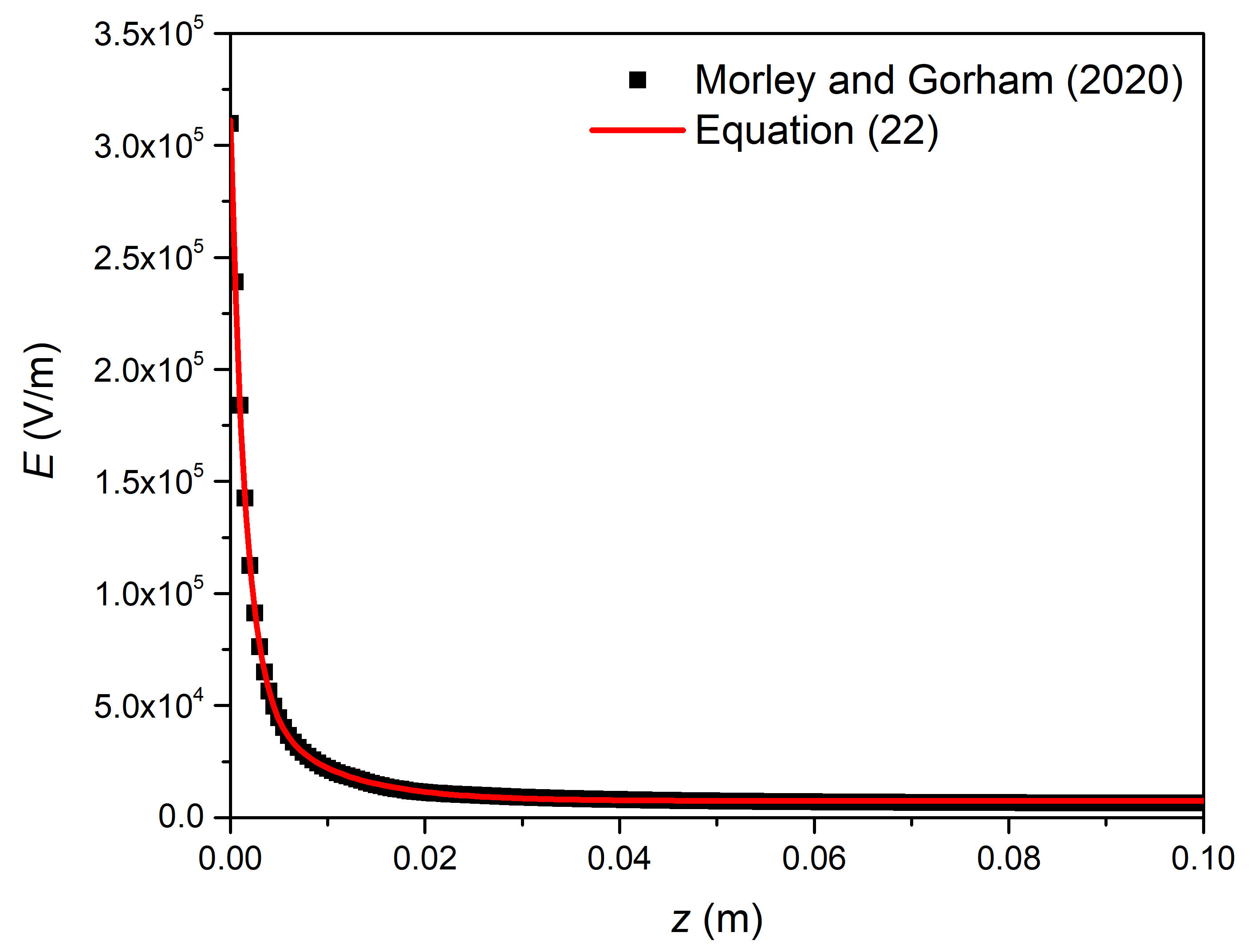}
\caption{Comparison of the electric potential obtained using equation.~\ref{eq::elecMorley} and that adopted in Morley and Gorham \cite{Morley2020}}
\label{fig:E_morley}
\end{figure}

It is worthy to note that Morley and Gorham~\cite{Morley2020} used an aluminum-foil covered prominence in order to concentrate the electric field near the tip. The electric potential obtained from equation~\ref{eq::elecMorley} is compared to that adopted by Morley and Gorham \cite{Morley2020} in Figure~\ref{fig:E_morley}. This figure shows a good agreement between the electric potential obtained from equation~\ref{eq::elecMorley} with that used by Morley and Gorham~\cite{Morley2020}. The Earth\rq{}s electrostatic field is on average much weaker than the one presented in equation~\ref{eq::elecMorley}; however, Morley and Gorham~\cite{Morley2020} pointed out the large variations in electrostatic field strength due to atmospheric activity that can generate the necessary lift for ballooning. The lift necessary for ballooning will be explored later in this paper (Figure~\ref{fig:plotVbarQbar}).

According to Morley and Gorham \cite{Morley2020}, the total required ballooning charge is around 1.28 nC/mg which is much smaller than that obtained by Gorham \cite{Gorham2013} due to higher electric potential field at the tip of the prominence that builds the electric field. In this simulation we consider a single 1
thread of length 0.5~m as in \cite{Morley2020}.

\begin{figure}[h!]
\centering
\includegraphics[width=0.5\textwidth]{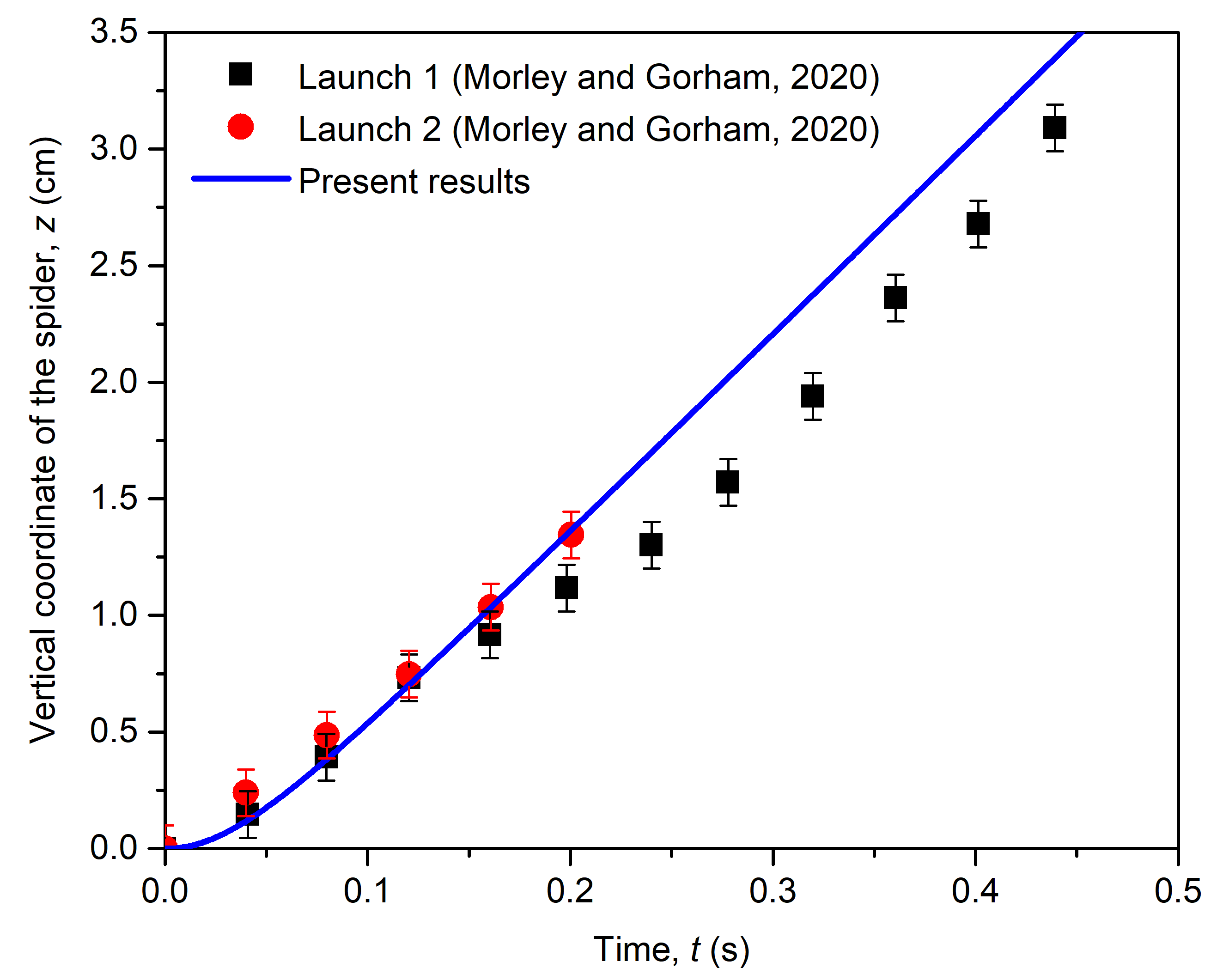}\\
(a)\\
\includegraphics[width=0.5\textwidth]{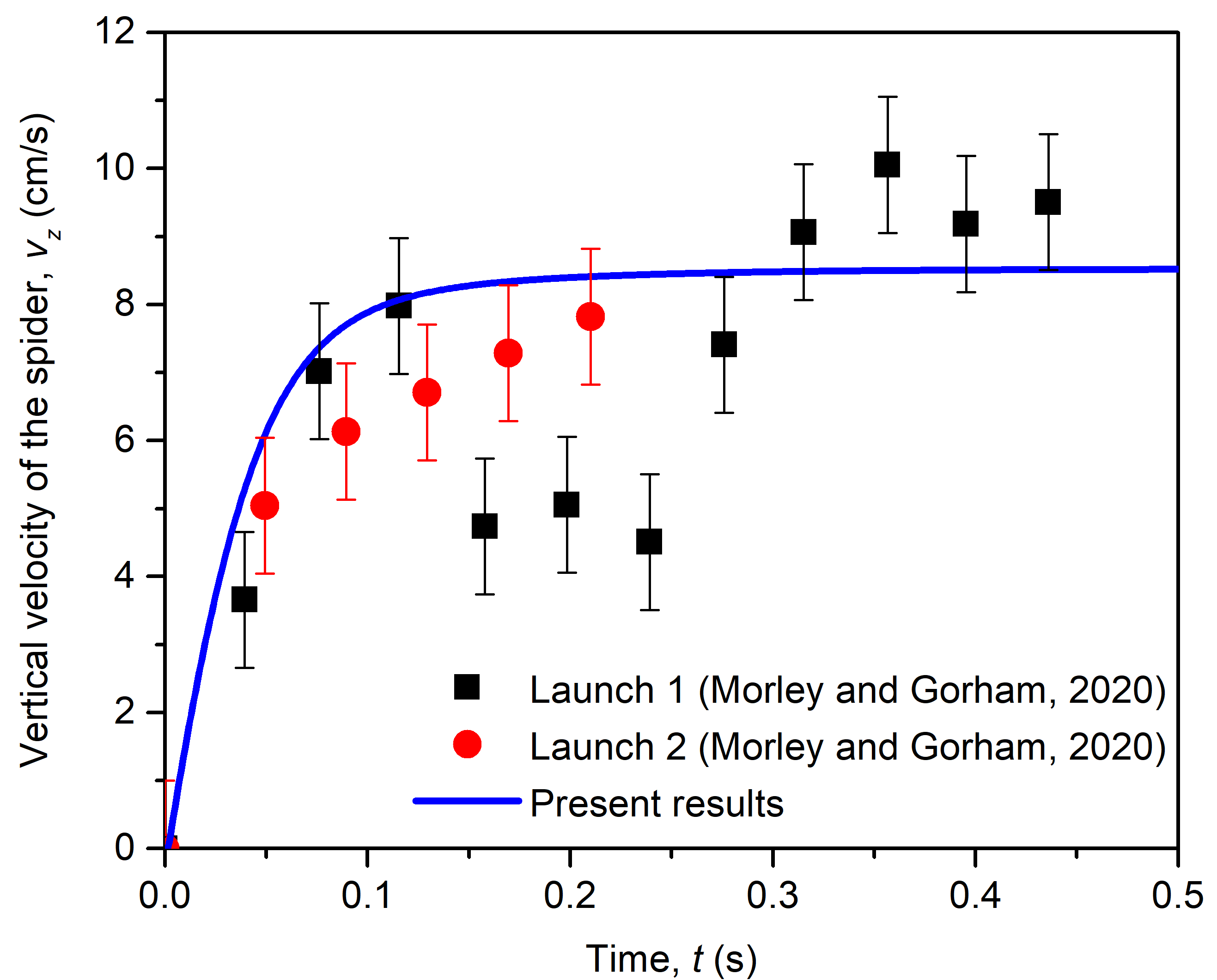}\\
(b)\\
\caption{(a) Vertical coordinate of the spider and (b) its vertical velocity compared to those obtained from Morley and Gorham \cite{Morley2020}}
\label{fig:ValidationMorley}
\end{figure}

Figure~\ref{fig:ValidationMorley} shows the comparison of the actual computed results for the vertical spider ballooning distance and the vertical spider ballooning velocity with those obtained experimentally by Morley and Gorham \cite{Morley2020}. In Figure~\ref{fig:ValidationMorley} (a), it is observed that the ballooning distance obtained in the present study corresponds well to that obtained experimentally. The comparison with the velocity in Figure~\ref{fig:ValidationMorley} (b) shows a fair agreement between the present computed results and those obtained experimentally by Morley and Gorham \cite{Morley2020}. It is observed that the spider is ejected promptly from the prominence where its velocity stabilizes at 8.5~cm/s after around 0.1 second.


\section{Results}
\label{sec::results}
In this section we present the results for ballooning velocity and thread unfolding dynamics. The electric field given in equation \ref{eq::elecMorley} is adopted.

\subsection{Normalized Quantities}
From principles of dimensional analysis, the ballooning phenomenon can be represented as a function of a number of non-dimensional (i.e. normalized) parameters. Hereon, we present our results in terms of normalized parameters. First, we introduce the relevant normalized quantities.

\begin{itemize}
\item
\textbf{Normalized terminal velocity.} Let us formulate a characteristic velocity, $v_\textrm{charac}$, from a scaling analysis based on the balance of forces along the $z$-axis. The force along the positive $z$-axis is $Q_t E_0 - W_s$, with $Q_t E_0$ the lift force and $W_s$ the spider weight (weight of the threads is negligible) and the viscous force along the negative $z$ scales as $\mu n_t l_t v_\textrm{charac}$. 
Balancing these forces, we get
\begin{equation*}
Q_t E_0 - W_s = \mu n_t l_t v_\textrm{charac};    
\end{equation*}
this leads to 
\begin{equation*}
v_\textrm{charac} = \frac{Q_t E_0 - W_s}{\mu l_t n_t}.    
\end{equation*}
The terminal velocity of the spider, $v_t$, is normalized by this characteristic velocity to obtain the normalized terminal velocity,
\begin{equation}
\overline{v}_t = \frac{v_t}{v_\textrm{charac}} = \frac{\mu v_t n_t l_t}{Q_t E_0 - W_s}
\label{eq:normV}
\end{equation}

\item
\textbf{Normalized viscous force.}
The viscous force scales as $F_v = \mu l_t^2 / t_\textrm{charac}$ and we use $t_\textrm{charac} = \sqrt{l_t/g}$ as the characteristic time. An estimate of the magnitude of the Coulomb repulsion force is $F_r = k_e Q_t^2 / [n_t^2 l_t^2]$. Normalizing the viscous force by the Coulomb repulsion force, we get the normalized viscous force,
\begin{equation}
\bar F_v = \frac{F_v}{F_r} = \frac{\mu n_t^2 \sqrt{g} l_t^{7/2}}{k_e Q_t^2}
\end{equation}

\item
\textbf{Normalized lift force.}
As sufficiently high altitude ($z \gg z_0$ in equation~\ref{eq::elecMorley}), the lift force scales as $Q_t E_0$. We normalize this by the weight of the spider to get the normalized lift force.
\begin{equation}
\overline{F}_l = \frac{Q_t E_0}{W_s}
\end{equation}

\item
\textbf{Normalized bending stiffness.} The characteristic bending force $YI/l_t^2$ is normalized by the characteristic Coulomb repulsion force $F_r = k_e Q_t^2 / [n_t^2 l_t^2]$ to get the normalized bending stiffness,
\begin{equation}
\overline{YI} = \frac{n_t^2 YI}{k_e Q_t^2}
\label{eq::normBend}
\end{equation}

\end{itemize}

The normalized lift force is varied by varying the total threads electric charge between 0.5 and 5~nC. The normalized bending stiffness is varied by varying the Young's modulus of elasticity $Y$ between 5 and 50~GPa where the average known silk modulus of elasticity is around 25~GPa \cite{Hudson2013}. The viscous force is varied by varying the viscosity between $10^{-7}$ and $10^{-3}$. The number of threads $n_t$ considered in this study are 1, 2, 4 and 8. Biologically, the number of threads observed in ballooning spiders range from 2 to 100 \cite{Wickler1986,Cho2018,Schneider2001}. However, to be able to explore the main physics of the spiders ballooning by parameter space exploration, and due to the associated computational limitations, the number of threads is limited to 8. Moreover, it is found that beyond 8 threads, there is no significant effect on the normalized terminal ballooning velocity of the spider as well as on the normalized lift and viscous forces.

\subsection{Ballooning Velocity}
\label{sec:ballooningSpeed}
In this section we present the variation of the normalized terminal ballooning velocity $\bar v_t$ in terms of normalized lift force $\bar F_l$, normalized viscous force $\bar F_v$ and normalized bending stiffness $\overline{YI}$ for different number of threads and for two cases:

\begin{itemize}
\item Electric charge located at the thread tip, as suggested by Morley and Gorham \cite{Morley2020}
\item Electric charge is uniformly distributed along the threads
\end{itemize}

\begin{figure}[ht]
\centering
\includegraphics[width=0.5\textwidth]{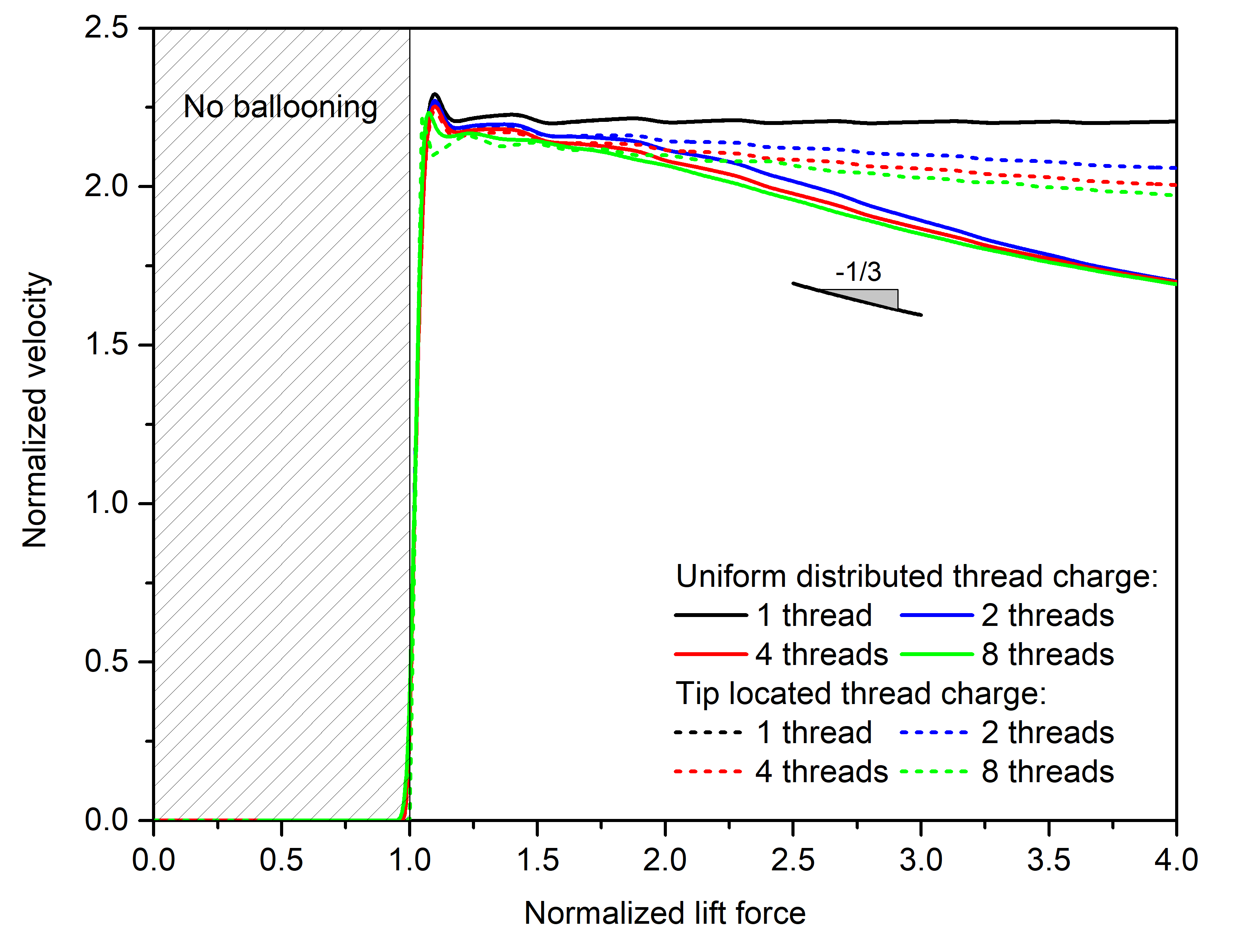}
\caption{Variation of the normalized terminal ballooning velocity versus normalized lift for a given value of normalized bending stiffness}
\label{fig:plotVbarQbar}
\end{figure}

In Figure~\ref{fig:plotVbarQbar} we compare the variation of $\bar v_t$ in terms of the normalized lift force for varying number of threads. This figure shows how the spider elicit ballooning once the normalized lift force exceeds 1. When the lift force is below 1, the ballooning velocity is zero since the spider cannot fly when the lift force generated by the electric potential field is smaller than the spider weight. For the tip located electric charge, the normalized terminal velocity slightly falls linearly with increasing normalized lift force. Referring to equation~\ref{eq:normV}, this implies that the dimensional velocity $v_t$ increases linearly with the dimensional lift force $Q_t E_0$.
On the other hand, for the uniform distributed thread charge the normalized velocity shifts from that for the tip located charge and decreases when the normalized lift force exceeds 1.5 with a slope around to $-1/3$, i.e. the dimensional velocity will increase with the dimensional lift force with a slope of approximately $2/3$ according to equation~(\ref{eq:normV}).
This dependence is caused by an intricate interplay between the Coulomb repulsion force, electrostatic lift force and the viscous drag. The repulsion force causes the threads to spread far apart from one another, i.e. all the threads would assume a horizontal configuration if the repulsion force was the only force acting on them. However, the shape of the thread influences the amount of viscous drag. According to the RFT, drag is the lowest when velocity is parallel to the tangent on the thread (i.e. the threads are vertical) and it is the highest when velocity is perpendicular to the tangent (i.e. the threads are horizontal). These competing forces cause the threads to deform to eventually find a configuration where all the forces sum to zero. Our simulation tool essentially solves this balance of forces and updates the configuration of the threads over time.

It is observed in this figure that for the one thread case and for $\bar F_l > 1$, the normalized terminal velocity stabilizes at around 2.2 while for the other cases with higher number of threads, the normalized terminal velocity reaches values around 2 for the tip located charge, and values around 1.7 for the uniformly distributed charge, especially for higher $\bar F_l$.
This further highlights the role of the deformed shape of the threads in common spider ballooning process. When there is only one thread, it is oriented vertically and thus experiences the least amount of drag. When multiple threads are introduced, the threads deform due to the Coulomb repulsion force and the threads are no longer oriented parallel to the direction of velocity. As such, more drag is exerted by air on the threads and the velocity is reduced in the case of still air. Meanwhile, the 3D conical thread net shape would lead to an increase in the ballooning speed in case of updraft wind caused by natural convection for instance.

In the present study we considered \textit{Erigon} spiders which are relatively small where their mass is about 1 mg and a size around 1 cm. Meanwhile, Schneider \textit{et al.} \cite{Schneider2001}, observed the ballooning of Stegodyphus dumicola (Eresidae) Pocock spiders weighting 100 mg and of size 7 to 14 mm. These spiders were found to balloon using 100 threads forming a triangular sheet with a length and width of about 1 m at the distal end. In the present study, we limited the number of silk threads to 8 due to computational limitations. However, referring to Figure~\ref{fig:plotVbarQbar}, we can see that the normalized terminal velocity becomes somehow independent from the number of threads when they exceed 8. And thanks to the normalized analysis, we can generalize our study to verify Schneider \textit{et al.} \cite{Schneider2001} observation regrading the ballooning of large spiders.

Using Figure~\ref{fig:plotVbarQbar} and data from Schneider \textit{et al.} \cite{Schneider2001}, we can deduce that if the spiders were on flat earth ground, where the electric field is 100 V/m, the electrostatic charge required for ballooning is around 100 nC per thread. From our simulations, it is observed that this very large electrostatic charge on the silk thread will lead to high Coulomb repelling forces which will cause the threads to repel diametrically in a plane which is in contradiction to Schneider \textit{et al.} \cite{Schneider2001} observations. Meanwhile, assuming the spiders are hanging on the top of tree branches where the electric field can reach 100 kV/m, the spider needs 0.1 nC per thread to balloon. It is worthy to note that Schneider \textit{et al.} \cite{Schneider2001} studies were done in farm Omdraai, Namibia which has very few trees. Moreover, the air temperature was reaching $33.8^{\circ}$ with almost no wind, a situation in favor for rising thermal currents. Thus, based on our conclusion and on Schneider \textit{et al.} \cite{Schneider2001} observations, for large spiders to balloon, rising thermal currents seem to be essential. In our study, we do not eliminate the fact that wind, turbulence and thermal currents could cause ballooning, however, we shed light on that these electrostatic forces could be alone used to balloon small spiders and that they are responsible on repelling the threads to avoid entanglement.

Referring to equation~(\ref{eq:normV}), and assuming that the following parameters are unchanged during typical ballooning: thread electric charge, $Q_t$, the atmospheric electric field, $E_0$, the air viscosity, $\mu$ and the spider weight, $W_s$, the spider could control its ballooning velocity by varying the thread length, $l_t$ and the number of threads. In the presence of significant wind speed, the spider could also control the flight altitude and direction by varying the number and length of ballooning threads. For instance, longer threads can result in larger drag forces and thus higher altitude in case of updrift wind. Reducing the length and number of threads could then be used during landing process.

The variation of the normalized terminal ballooning velocity versus normalized viscous force for $Q_t = 2.5~\mathrm{nC}$ and $Y = 20\times10^{9}~\mathrm{Pa}$ was also analyzed. It is shown that the normalized terminal ballooning velocity is always equal approximately to 2.12. This indicates that the normalized viscous forces do not play a major role in the ballooning of spiders once it reaches steady-state.

We also observed in our simulations that the twisting and stretching deformation is negligible compared with bending mode. In summary, once the spider reaches a steady velocity after the transient dynamics, the velocity and the shape of the threads do not depend on the normalized viscous forces.

The variation of the normalized ballooning velocity versus normalized bending stiffness for $Q_t = 2.5~\mathrm{nC}$ and $\mu =18.37\times10^{-6}~\mathrm{Pa.s}$ shows also that the normalized terminal ballooning velocity is always equal to approximately 2.12. This indicates that the bending stiffness, representative of the elasticity of the thread, does not play a major role in the ballooning of spiders. This is also anticipated from the ratio of characteristic bending force to characteristic Coulomb repulsion force. This ratio is defined as the normalized bending stiffness in equation~\ref{eq::normBend} and its value in the regime relevant to ballooning is on the order of $10^{-10}$. We also observed in our simulations that the twisting and stretching deformation is negligible compared with bending mode. In summary, once the spider reaches a steady velocity after the transient dynamics, the velocity and the shape of the threads do not depend on the elastic bending, stretching, and twisting stiffness.

\subsection{Thread unfolding Dynamics}
\begin{figure*}[ht]
\begin{subfigure}[b]{0.1\textheight}
\centering
\includegraphics[height=0.15\textheight]{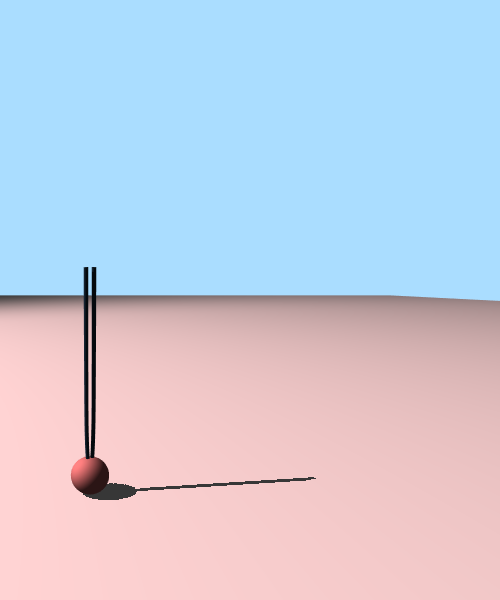}
\caption{$7.22 \times 10^{-3}~\mathrm{s}$}
\end{subfigure}
\hfill
\begin{subfigure}[b]{0.1\textheight}
\centering
\includegraphics[height=0.15\textheight]{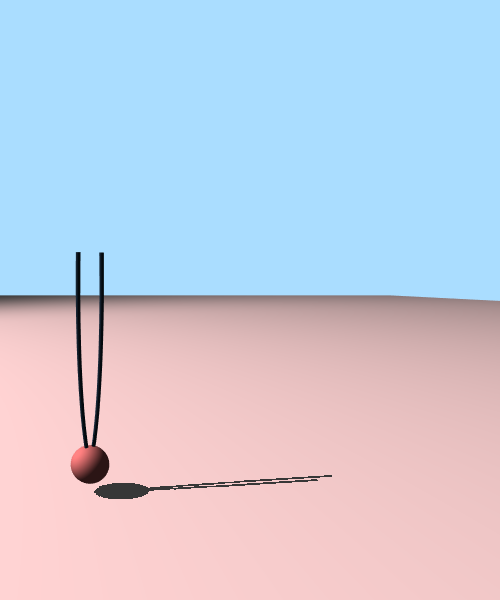}
\caption{$7.22 \times 10^{-2}~\mathrm{s}$}
\end{subfigure}
\hfill
\begin{subfigure}[b]{0.1\textheight}
\centering
\includegraphics[height=0.15\textheight]{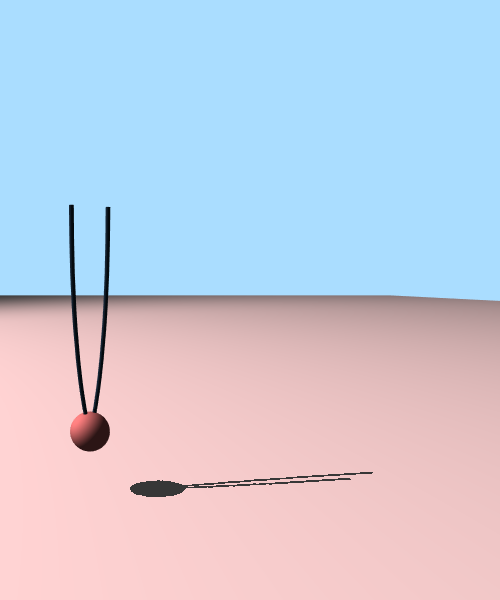}
\caption{$1.72 \times 10^{-1}~\mathrm{s}$}
\end{subfigure}
\hfill
\begin{subfigure}[b]{0.1\textheight}
\centering
\includegraphics[height=0.15\textheight]{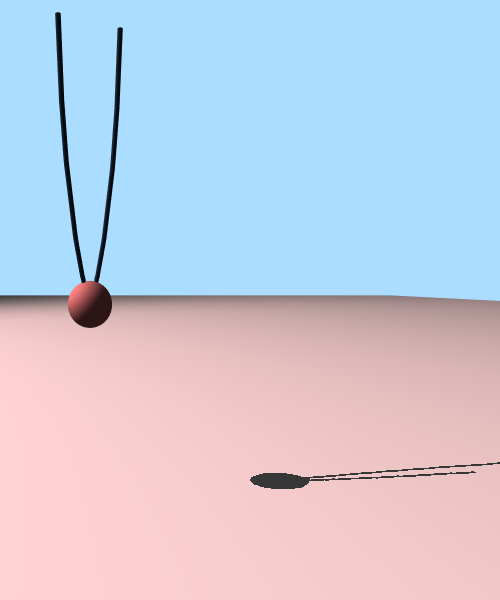}
\caption{$4.22 \times 10^{-1}~\mathrm{s}$}
\end{subfigure}
\newline
\begin{subfigure}[b]{0.1\textheight}
\centering
\includegraphics[height=0.15\textheight]{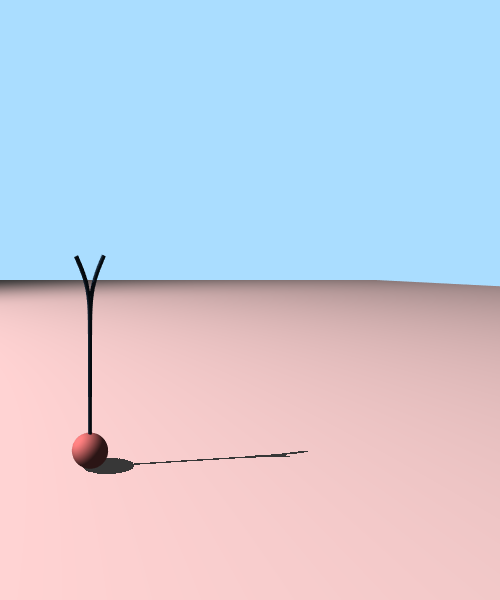}
\caption{$2.02 \times 10^{-3}~\mathrm{s}$}
\end{subfigure}
\hfill
\begin{subfigure}[b]{0.1\textheight}
\centering
\includegraphics[height=0.15\textheight]{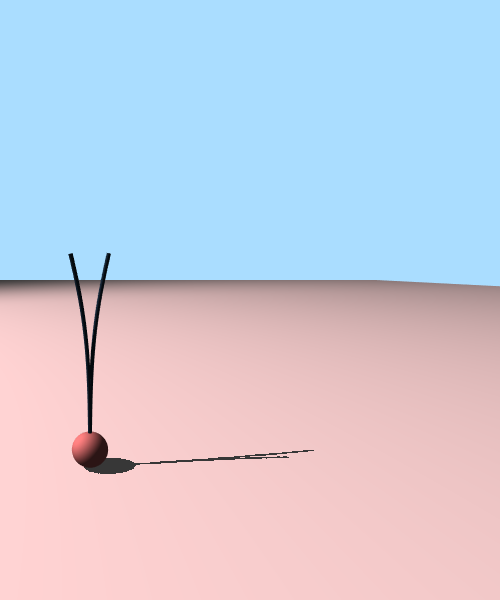}
\caption{$2.22 \times 10^{-2}~\mathrm{s}$}
\end{subfigure}
\hfill
\begin{subfigure}[b]{0.1\textheight}
\centering
\includegraphics[height=0.15\textheight]{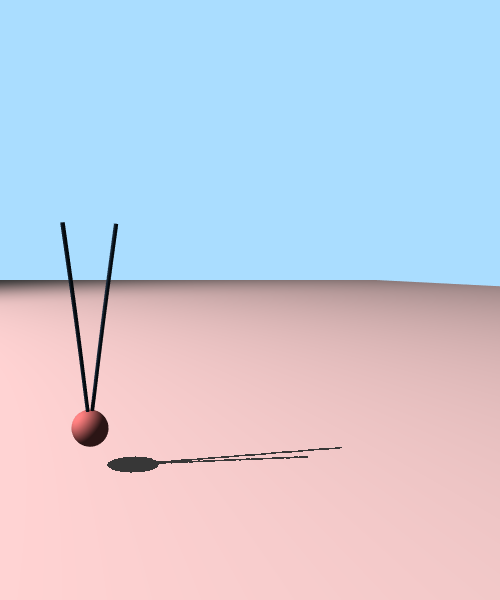}
\caption{$1.22 \times 10^{-1}~\mathrm{s}$}
\end{subfigure}
\hfill
\begin{subfigure}[b]{0.1\textheight}
\centering
\includegraphics[height=0.15\textheight]{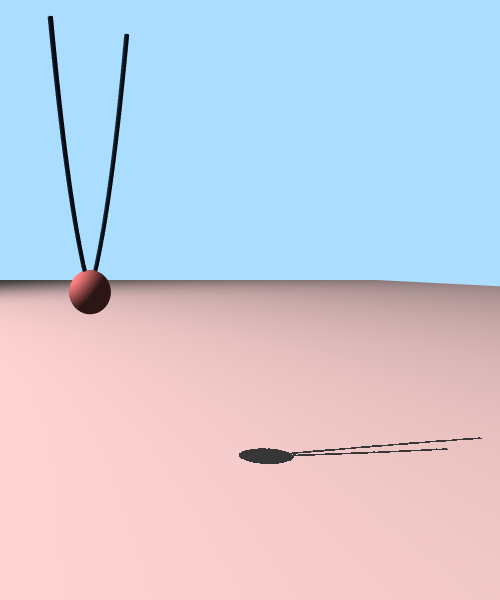}
\caption{$4.22 \times 10^{-1}~\mathrm{s}$}
\end{subfigure}

\caption{Unfolding dynamics of two spider threads with uniformly distributed (a-d) and tip located (e-h) electric charge}
\label{fig:bendingDynamics}
\end{figure*}

The spider threads unfolding dynamics is shown in Figure~\ref{fig:bendingDynamics} for the 2 threads case for better visibility. For the uniformly distributed thread charge it can be observed that the bending occurs along the threads which move apart gradually with an increase in their curvature until they reach a steady-state position with a v-shape as observed in real ballooning spiders.

For the case with tip located charge, the repelling force acts on the tips of the threads pushing them apart while maintained in close contact in the bottom region. After a short time, the threads make a v-shape similar to that of the previous case. The results are also accompanied with animations showing the time evolution of the unfolding dynamics (see Supplementary Material).

\begin{figure*}[ht]
\centering
\includegraphics[width=0.9\textwidth]{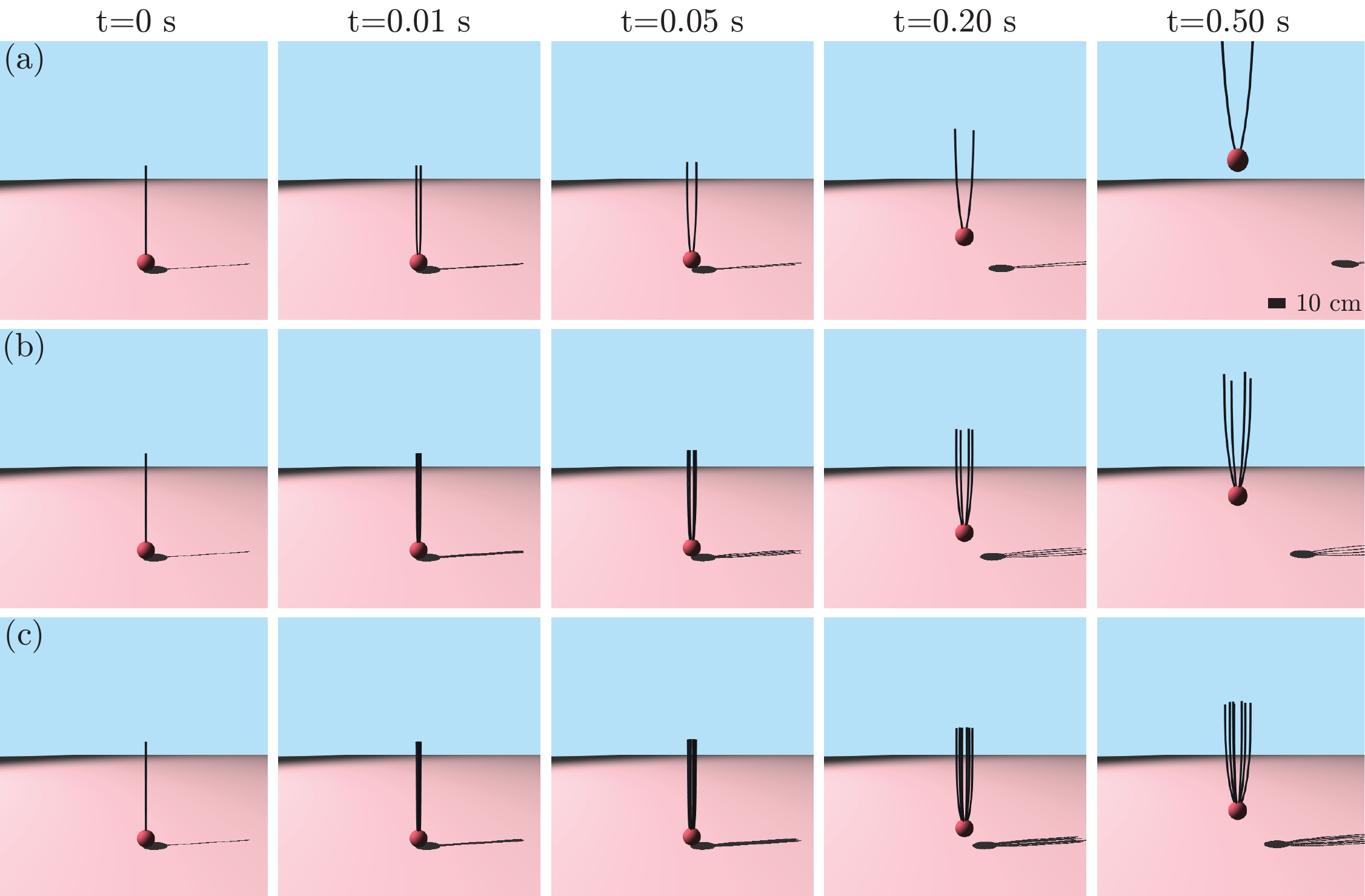}
\caption{Ballooning process obtained from our 3D numerical simulations for spider with (a) two threads, (b) four threads and (c) eight threads. A supplemental animation is also available at \url{https://youtu.be/RKnIp5ERVbE}.}
\label{fig:introFig}
\end{figure*}

Figure~\ref{fig:introFig} shows the ballooning process and spider threads bending for two, four and eight threads. The threads are pushed apart due to the Coulomb electrostatic forces while the spider is moving upward due to the atmospheric electric field. A supplemental material also shows an animation of the multi-threaded spider ballooning and unfolding dynamics.

\section{Conclusion}
\label{sec::conclusion}
Three dimensional numerical simulations are performed for spider ballooning due to electrostatic forces. Spiders with multi silk threads are considered in this study. The numerical method of the fluid-structure-electric field interaction combines the DER algorithm to compute the elastic deformation of the spider threads. Moreover, the RFT is used to compute the hydrodynamic viscous forces on the spider and on the threads. The electrostatic forces caused by the atmospheric potential gradient and the thread electric charge is computed based on the Coulomb theory. The spider is approximated by a sphere attached to one or multiple silk threads. The numerical results computed in this paper are first validated against theoretical and experimental data from the open literature for one-thread case showing a good agreement.

Two cases were studied in this paper. In the first one, we assume that the thread charge is uniformly distributed along the threads. In the second, we assume that the charge is located at the thread tip.

The results show that for one thread case, the normalized velocity is around 2.2 and independent of the normalized lift and normalized viscous forces while it is slightly less for the multi-thread cases. In the uniformly distributed charge case, the normalized ballooning velocity deviates from that for the tip located charge and decreases slightly when the normalized lift force exceeds 2 with a slope equal to -1/3.

Finally, the Coulomb repelling forces cause the threads to bend and form a three-dimensional conical sheet very similar to observations from open literature. This bending behavior is very fast and occurs in the beginning of the ballooning process before it stabilizes at a steady-state shape.

It should be noted that the wind speed and its fluctuations could affect the behavior of trichobothria. In fact, the spiders use the deformation of trichobothria signal to determine whether they will balloon or not. Hence, for high wind speeds and fluctuations the signals of electric field could be burried and the spider may not be able to distinguish whether the deformation of trichobothria is caused by wind or by electric field. Therefore, spiders usually balloon on relatively calm days as explained earlier in the introduction section.

Moreover, in this study, the aim is to explore the essential physics of the spiders ballooning by parameter space exploration and therefore a computationally efficient framework with DER and RFT has been chosen. A more comprehensive model coupling for instance SBT and DER is an interesting direction for future research.

\begin{acknowledgments}
The authors would like to thank E.L. Morley from the University of Bristol and P.W. Gorham from the University of Hawaii at Manoa for providing the electric field data in the chamber where the experiments were done in reference \cite{Morley2020} and used in the present paper to generate the electrostatic field model given in equation~\ref{eq::elecMorley}. M.K.J. acknowledges support from the National Science Foundation (Award numbers: CAREER-$2047663$, CMMI-$2053971$, CMMI-$2101751$, IIS-$1925360$).
\end{acknowledgments}

\clearpage

\bibliography{biblio}

\end{document}